\documentstyle[pra,aps]{revtex}
\begin{document}

\title{
Mechanism for Magnetic Flux Generation in
Grain Boundaries of YBa$_2$Cu$_3$0$_{7-x}$}

\draft
\author{M. B. Walker}
\address{Department of Physics, University of Toronto,
Toronto, Ontario, M5S 1A7, Canada}
\date{\today}
\maketitle 


\begin{abstract}
The spontaneously generated magnetic flux observed by Mannhart
{\it et.\ al.\ } (1996) in asymmetric $45^{\circ}$ grain boundaries in 
YBa$_2$Cu$_3$O$_{7-x}$
was explained by them in terms of a $d_{x^2-y^2}$ superconducting order
parameter and grain boundary faceting.  This article argues that
twin boundaries which contact the grain boundary also play an
important role.
\end{abstract}


\section*{Introduction}

Recently Mannhart {\it et.\ al.\ }\protect\cite{man96} 
and Moler {\it et.\ al.\ }\protect\cite{mol96} 
have observed spontaneously generated
delocalized magnetic flux in what they call asymmetric $45^{\circ}$
grain boundaries in YBa$_2$Cu$_3$0$_{7-x}$ (YBCO). They have explained the 
existence
of this flux in terms of a superconducting order parameter having
$d_{x^2-y^2}$ symmetry and grain boundaries which are faceted. The
geometry of their experiment and of their explanation is indicated in
Fig.~1.  In the words of Ref.~\protect\cite{mol96}, ``For a
$d$-wave
superconductor, a lobe of the order parameter is normal to the
boundary on one side, while a node is normal on the other. 
Faceting of the grain boundary produces slight variations of the
grain boundary angle, rocking the local normal from the positive
lobe to the negative lobe, varying the local critical current from
positive to negative, producing a series of 0 and $\pi$-junctions
and spontaneous magnetization.''  Ref.~\protect\cite{man96} notes
that
the variation of the magnetic flux along the grain boundary is
described by the equation~\protect\cite{fer63,jos65}
\begin{equation}
{\lambda_J}^2{d^2 \gamma (y)\over dy^2} 
=\sin[\gamma (y)-\alpha(y)]
\label{eq1}
\end{equation}
where $y$ is the distance measured along the grain boundary, 
$\gamma (y)$
is the gauge-invariant phase difference across the boundary at
point $y$, and $\lambda_J$ is the Josephson penetration depth. The
idea that the variations of the grain boundary angle cause the
local critical current to vary from positive to negative is
incorporated into this equation by taking $\alpha(y)$ to be zero
when the critical current is positive, and $\pi$ where the critical
current is negative. The flux in the grain boundary, integrated
from one end of the grain boundary up to the point $y$, is given by
$\Phi(y) =\Phi_0\gamma (y)/2\pi$ where the $\Phi_0$ is the elementary
flux quantum.  In a realistic model of a grain boundary,
$\lambda_J$ would be a function of $y$ (as in Ref.~
\protect\cite{man96}) but this dependence is neglected in the
present article.

The purpose of this article is to present an alternative physical
model which also leads to spontaneous flux generation in
asymmetric 45$^\circ$ twin boundaries, and which could
thus also account for the experimentally observed behavior. In this
alternative model, which also leads to Eq.~\protect\ref{eq1},
the changes of $\alpha(y)$ from 0 to $\pi$ occur
where twin boundaries from one side of the grain boundary contact
the grain boundary (twin boundaries on the other side of the grain
boundary do not cause $\alpha(y)$ to change). The details of how
this twin-boundary mechanism works depend on the fact that YBCO is
orthorhombic, and also on the assumption that the twin boundaries
in YBCO have odd reflection symmetry ({\it e.g.} see 
Refs.~\protect\cite{wal96a,wal96b}). The essential idea is that,
given these assumptions, the point $y=0$ of 
Fig.~\protect\ref{fig2}(a) can be viewed as a meeting point of
three Josephson junctions, an odd number of which must be $\pi$ 
junctions.  Under these circumstances, it is known from the
work of Refs.~\protect\cite{ges87,kir96,tsu96} that the superconducting
ground state will have a vortex containing a flux
${1\over 2}\Phi_0$ at the meeting point of the junctions.

\section*{Twin boundary mechanism}

Consider the situation shown in Fig.~\protect\ref{fig2}(a).  On the
left side of the grain boundary there are two distinct twins
separated by a twin boundary and characterized by their 
Ginzburg-Landau order parameters $\psi_1$ and $\psi_2$,
respectively; the superconductor on the right side of the grain
boundary is characterized by the order parameter $\psi_0$.  In a
Ginzburg-Landau model for the grain boundary, the relevant
contribution to the superconducting free energy per unit area of
the grain boundary is
\begin{equation}
f_> = C Re (\psi_2\psi^\ast_0)
\label{eq2}
\end{equation}
for points $y$ on the positive $y$ axis, and
\begin{equation}
f_< = D Re(\psi_1\psi_0^\ast)
\label{eq3}
\end{equation}
for points on the negative $y$ axis ({\it e.\ g.}\ see
Ref.~\protect\cite{and87}).  Now consider a new state obtained by
reflecting the sample of Fig.~\protect\ref{fig2}(a) in a plane
normal to the plane of the figure and containing the twin boundary.
Assuming that the YBCO superconducting order parameter is of the
$A_{1g}$ (or $ux^2 + vy^2)$ orthorhombic symmetry type ({\it e.\
g.}\ see Refs.~\protect\cite{wal96a,wal96b}), the new order
parameters are related to the old by $\psi_1^{new} = \psi_2,
\psi_2^{new} =\psi_1$ and $\psi_0^{new}=\psi_0$.  Since the total
free energy is unchanged by such a reflection, $C=D$ in
Eqs.~\protect\ref{eq2} and \protect\ref{eq3}.

The interpretation
\protect\cite{wal96a,wal96b} of Josephson experiments
\protect\cite{wol93,mat95,wol95,tsu94} on twinned crystals has
shown that the superconducting state of YBCO has odd symmetry
with respect to a reflection in a twin boundary, {\it i.\ e.}\ that
$\psi_2=-\psi_1$.  The argument is as follows.  In
Ref.~\protect\cite{and87} it is shown that the superconducting
order parameter can either be continous or change sign at a
twin boundary; for YBCO these two cases correspond to the
superconducting state having either even or odd symmetry
with respect to a reflection in the twin boundary.  Now,
Josephson experiments on twinned crystals determine what is
called the macroscopic symmetry of the superconducting
state of the twinned crystal (which
is different from the microscopic symmetry of the superconducting
state of a single twin).  It can be shown that, if the twin
boundaries have even reflection symmetry, the macroscopic
symmetry of the superconducting state of the twinned crystal
is that of a tetragonal s-wave superconductor.  On the other
hand, if the twin boundaries have odd reflection symmetry, the
macroscopic symmetry is that of a tetragonal d$_{x^2-y^2}$ 
superconductor.  Since the Josephson experiments on twinned
crystals \protect\cite{wol93,mat95,wol95,tsu94} show
macroscopic d$_{x^2-y^2}$ symmetry, the superconducting
twin boundaries must have odd reflection symmetry. Further
details of this argument can be found in 
\protect\cite{wal96a,wal96b}.

A twin boundary at which the superconducting state has odd
reflection symmetry (i.e. for which $\psi_1=-\psi_2$) can be
viewed as a $\pi$ Josephson junction (a $\pi$ junction because
of the change of sign of the order parameter there) with a very high 
critical current density. Because of the very high value of 
the critical current density (which is much greater than the current 
densities occuring in the superconductor)
the phase difference of the order parameter across the junction is
always locked to $\pi$.

Now note that the point $y=0$ in Fig.~\protect\ref{fig2}(a)
can be viewed as the meeting point of three Josephson
junctions.  Because $C=D$ in
Eqs.~\protect\ref{eq2} and \protect\ref{eq3} the grain boundary
Josephson junctions along $y>0$ and $y<0$ are either both
normal junctions or both $\pi$ junctions.  Also, as discussed above, 
since $\psi_2 = - \psi_1$ along the twin boundary, the twin
boundary can be viewed as a $\pi$ junction.  Clearly, an odd number of
$\pi$ junctions converge at the meeting point.
Thus, the stable superconducting
ground state will contain a vortex with flux ${1\over 2}\Phi_0$
at the point $y=0$ in agreement with the arguments and
experimental observations of \protect\cite{kir96} and
\protect\cite{tsu96}.  

As noted in Ref.~\protect\cite{sig92}, the designation of a
Josephson junction as a normal junction or a $\pi$ junction
is a matter of convention (the total number of $\pi$ junctions
which meet at a point is however fixed).  In the above, the
given transformation properties of the order parameter under a 
reflection in the twin boundary are what imposed the particular
convention adopted. If the substitution 
$\psi_2 = -\tilde{\psi}_2$ is made, then the twin boundary
will become a normal junction and the grain boundary junction 
for $y > 0$ will become a $\pi$ junction.  Note that this
substitution produces an extra minus sign in 
Eq.~\protect\ref{eq2}; it is this additional minus
sign that leads to $\alpha (y) = \pi$ for $y>0$ in
Eq.~\protect\ref{eq1}, while $\alpha (y) = 0$ for
$y<0$.

It is also important to note that twin boundaries on the other side
of the grain boundary, as in Fig.~\protect\ref{fig2}(b), are not
expected to lead to variations of $\alpha(y)$.  As above, the
superconducting free energy per unit area of the grain boundary
above and below the point zero can be written
\begin{equation}
f_>=A Re(\psi_2\psi_0^{\ast})\ ,
\label{eq4}
\end{equation}
and
\begin{equation}
f_<= B Re(\psi_1\psi_0^{\ast})\ .
\label{eq5}
\end{equation}
The assumption that the superconductivity of YBCO is predominantly
of the type $d_{x^2-y^2}$ implies that $A$ and $B$ have opposite signs. 
In addition, for the reasons stated above, $\psi_2 =-\psi_1$.  The
combination of these two sign changes means that there will be no
effect on the net contribution to $\alpha(y)$ when crossing a twin
boundary of the type indicated in Fig.~\protect\ref{fig2}(b).

The arguments just given show that the superconducting grain
boundary free energy can be written
\begin{equation}
F=-{\Phi_0\over 2\pi c} j_c L\int dy\left[{1\over
2}{\lambda_J}^2\left({d\gamma \over dy}\right)^2 -V(\gamma)\right]
\label{eq6}
\end{equation}
where
\begin{equation}
V(\gamma) =\cos [\gamma(y)-\alpha(y)]\ ,
\label{eq7}
\end{equation}
$j_c$ is the critical current density, $L$ is the thickness of the
superconducting film in the direction normal to the plane of
Fig.~\protect\ref{fig2}, and $c$ is the speed of light.  Also,
${\lambda_J}^2 =\Phi_0c/(8\pi^2 dj_c)$ where $d\approx 2\lambda_L$
is the magnetic thickness of the grain boundary, $\lambda_L$ being
the London penetration depth. The quantity proportional to
$\lambda_J^2$ represents the magnetic and surface current energy in
the grain boundary.\protect\cite{jos65}

The free energy of Eq.~(\protect\ref{eq6}) has the form of a
Lagrangian describing the motion of a particle of mass
${\lambda_J}^2$ in a potential $V(\gamma )$; here $\gamma$ represents the
particle displacement, and $y$ is time.  Thus, it is clear that the
condition for $F$ to be a minimum is the equation of motion (1),
where the force on the particle is $-\partial V/\partial
\gamma =\sin(\gamma -\alpha)$.  Since the particle velocity, $d\gamma /dy$, is
proportional to the magnetic induction $h(y)$ in the junction, {\it
i.\ e.}\ 
\begin{equation}
{d\gamma \over dy}= 4\pi\lambda_Lh(y)/\Phi_0\ ,
\label{eq8}
\end{equation}
it will be continuous across the twin boundaries where $\alpha$
changes by $\pi$; also, since $d\gamma /dy$ is continuous, 
$\gamma $ will also
be continuous.  This interpretation of Eq.~(\protect\ref{eq1}) as
describing the motion of a particle in a potential $V(\gamma )$ allow a
simple intuitive description of the solutions of
Eq.~(\protect\ref{eq1}).  

\section*{Josephson vortices}

First consider the solution of Eq.~\protect\ref{eq1} for
$\alpha=0$, {\it i.\ e.}\ no twin boundaries present.
\protect\cite{fer63,jos65}  There will be a solution corresponding
to a particle starting at the potential maximum (point $a$ of
Fig.~\protect\ref{fig3}(a)) with zero velocity, accelerating down
the hill to reach its maximum velocity at point $b$, and then
gradually decelerating as it climbs the hill to come to rest again
at point $c$. The corresponding graph of displacement versus time
(or $\gamma $ versus $y$) is shown in Fig.~\protect\ref{fig3}(b).  The
total change in $\gamma $ of $2\pi$ gives a flux associated with this
vortex of $\Phi_0$ (recall $\Phi(y)=\Phi_0 \gamma (y)/(2\pi)$).  Since
the magnetic induction in the grain boundary is proportional to the
particle velocity $d\gamma /dy$, it has its maximum value at the point
$y=0$ of Fig.~\protect\ref{fig3}(b).  For this example, the solution
$\gamma (y)$ and the free energy can be evaluated
analytically.\protect\cite{jos65}  The free energy of this vortex
(relative to the state with $\gamma (y)= {\rm constant} = 0$) is
positive, which means that the vortex is not stable in the absence of an
external magnetic field.

\section*{Vortex with flux \protect{${1\over 2}\Phi_0$} trapped
by a twin boundary}

Now consider the case of Fig.~\protect\ref{fig2}(a) where the grain
boundary is contacted by a single twin boundary at $y=0$, and there
are no other twin boundaries within a distance of many Josephson
penetration depths $\lambda_J$.  Here, $\alpha=0$ for $y<0$ and
$\alpha=\pi$ for $y>0$. The relevant potential is shown in
Fig.~\protect\ref{fig4}(a), and the solution of interest
corresponds to a particle starting at point $a$ with zero velocity,
accelerating down the hill to point $b$, decelerating up the hill
from $b$ and coming to rest at $c$. As noted above, the particle
velocity is continuous at point $b$. The corresponding graph of $\gamma $
as a function of $y$ is shown in Fig.~\protect\ref{fig4}(b). The
change of $\gamma $ by $\pi$ corresponds to the vortex containing a flux
of ${1\over 2}\Phi_0$. This vortex is confined to be centred on the
twin boundary, since moving the vortex away from the twin boundary
will increase its free energy by an amount which, at large
distances, is proportional to the distance it is displaced. It is
also clear that the state with no vortex ({\it i.\ e.}\ $\gamma ={\rm
constant}$) is unstable with respect to the formation of the
${1\over 2}\Phi_0$ vortex state.

A related problem is that in which there are an odd number of twin
boundaries like the one in Fig.~\protect\ref{fig2}(a) which contact
the grain boundary in a region $y_1<y<y_2$, but where there are no
changes of $\alpha$ ({\it i.\ e.}\ no twin boundaries) for several
Josephson penetration depths for $y>y_2$ and $y<y_1$.  In this case
also, the state with no net flux in the region $y_1<y<y_2$ will
be unstable with respect to the formation of a state with a flux of
${1\over 2}\Phi_0$ (states with flux $(n+{1\over 2})\Phi_0$, $n$ an
integer, are also possible).

\section*{Pinning of a Josephson vortex}

Suppose the grain boundary contains only two twin boundaries of the
type shown in Fig.~\protect\ref{fig2}(a), and that these are
separated by a distance $d\ll \lambda_J$. The appropriate potential
is shown in Fig.~\protect\ref{fig5}.  Again the particle begins at
$a$ with zero velocity, accelerates downhill, passes the points $b$
and $c$ corresponding to the location of the two twin boundaries,
and eventually comes to rest at $d$.  In passing point $b$, the
phase $\alpha$ jumps by $\pi$; because $d\gamma /dy$ must be continuous
at $b$, the kinetic energy is continuous at $b$; the potential
energy of the particle in the region from $b$ to $c$ is therefore
taken to be $C+\cos[\gamma (y)+\pi]$ where the constant $C$ is chosen so
that the potential energy, and hence the total energy, will be
conserved when the particle passes point $b$.  Because of the time
the particle spends in the region from $b$ to $c$, its free energy
is lowered relative to what it would be in the absence of twin
boundaries by an amount which is approximately $\Phi_0 j_c Ld/(\pi
c)$. This means that the vortex will be pinned by the region
between the two twin boundaries.

\section*{Random, closely-spaced twin boundaries}

Suppose the separation between twin boundaries is random and
typically has a magnitude less than the Josephson penetration depth
$\lambda_J$.  Mannhart {\it et.\ al.}\protect\cite{man96} have
discussed a stochastic approach to the estimation of the variations
of $\gamma (y)$ in this case, and have shown how a flux of $\Phi_0$ can
be spread out over a large distance ({\it i.\ e.} a distance of
many times $\lambda_J$) as is observed in their experiments.  This
spreading out of the flux can also be understood in terms of the
dynamical picture discussed above. Consider, for example, a
particle beginning with zero velocity at $a$ in the potential of
Fig.~\protect\ref{fig6} and accelerating downhill to $b$ where it
encounters a twin boundary; the particle then climbs from $b$ to
$c$; at $c$ the particle velocity goes to zero (the potential
energy at $c$ is the same as at $a$) and the particle then rolls
back downhill past $b$, $c$ and comes to $f$, where another twin
boundary is encountered, and then continues towards $g$. It is seen
that the particle can spend a lot of time rolling back and forward
in the various potentials it encounters each time a twin boundary
is crossed. This time is equivalent to distance measured along the
twin boundary. This shows how a small amount of total flux can be
spread out over a large distance.

\section*{Conclusion}

The fundamental process which is at the origin of spontaneous flux
generation in asymmetric $45^{\circ}$ grain boundaries in the
mechanism advanced in this paper is the stabilization of a vortex
with flux ${1\over 2}\Phi_0$ at points where a twin boundary such
as that shown in Fig.\protect\ref{fig2}(b) intersects a grain
boundary. The existence of such half integral vortices is a
consequence of the fact that the superconducting state of YBCO has
odd symmetry with respect to a reflection in a twin boundary, and
the observation of such vortices would thus constitute an
experimental test of this odd reflection symmetry. Furthermore,
this conclusion is independent of whether the $A_{1g}$
orthorhombic-symmetry order parameter of YBCO is predominantly
$d_{x^2-y^2}$ or s-wave.

In contrast to the above, the point where the twin boundary of
Fig.~\protect\ref{fig2}(b) intersects the grain boundary should
not support the existence of a vortex containing a flux of ${1\over
2}\Phi_0$. In this case, the assumption that the YBCO
superconductivity is predominantly $d_{x^2-y^2}$-wave has been combined
with the assumption that the twin boundaries have odd reflection
symmetry to arrive at this conclusion.

The fundamental process which is at the origin of spontaneous flux
generation in the mechanism of Mannhart {\it et.\
al.}\protect\cite{man96} and Moler {\it et.\
al.}\protect\cite{mol96} is the stabilization of a vortex with a
flux of ${1\over 2}\Phi_0$ at the vertex $b$ 
in Fig.~\protect\ref{fig1} of the intersection of
two distinct facets of a grain boundary. It is possible that both
the faceting mechanism and the twin boundary mechanism play a role
in grain boundaries studied in Ref.~\protect\cite{man96} and
\protect\cite{mol96}.

Although the materials and sample preparation problems are no doubt
formidable, it would be of interest to prepare samples having
isolated twin boundaries, and an isolated intersection of two
facets [such as shown in Figs~\protect\ref{fig1} and
\protect\ref{fig2}(a) and (b)] and to observe directly the presence
or absence of vortices with a flux of ${1\over 2}\Phi_0$ in these
idealized cases.
The superconductor on the right hand side of
Figs.~\protect\ref{fig1} and \protect\ref{fig2}(a), and on the left
of Fig.~\protect\ref{fig2}(b), does not necessarily have to be
YBCO, but could be an isotropic superconductor such as lead.

An interesting special case of the faceting mechanism for 
the stablization of a
${1\over 2}\Phi_0$ vortex occurs in the corner junction such
as that used in Ref.~\protect\cite{wol95}.  To obtain the
${1\over 2}\Phi_0$ vortex, the junction would have to be long
in comparison with the Josephson penetration depth. The actual
experimental conditions of Ref.~\protect\cite{wol95}
correspond to the short junction case.

Note also that in c-axis tunneling from a
twinned crystal of YBCO to Pb (as in Ref. \protect\cite{sun94})
the Josephson junctions formed by the two different types of twins
must be, one a $\pi$ junction and the other a normal 
junction (see Ref.\protect\cite{wal96a}).
Thus, a necessary condition for spontaneous flux generation is
satisfied here also. For example, for a YBCO crystal which 
consists of only two twins, a vortex containing a half a flux
quantum would form along the line where the plane containing the
single twin boundary contacts the junction.  The fact that the
junctions\protect\cite{sun94} are ``fabricated on a broad surface
normal to the $c$ axis'' however means that the normal and $\pi$
junctions are distributed over a two-dimensional surface.  This is 
different from the situation of Refs. \protect\cite{man96,mol96} 
where the normal and $\pi$ junctions are distributed
along a one-dimensional grain boundary.

The degree of orthorhombicity of the superconductivity of YBCO
plays a role in the mechanisms just described. For example, for a
tetragonal $d_{x^2-y^2}$\ superconductor, the critical current for an
ideal asymmetric $45^{\circ}$ twin boundary (such as $oo^{\prime}$
in Fig.~\protect\ref{fig1}) is zero, and the critical currents on the grain 
boundary facets $ab$ and $bc$ have opposite signs.  For a superconductor which 
 has some orthorhombicity but is nevertheless approximately $d_{x^2 - y^2}$, 
the critical current on $oo^{\prime}$ of Fig.~\protect\ref{fig1} is not zero,
and the angle $\delta$ (or
$\delta^{\prime}$) will have to exceed some nonzero critical value
in order for the critical currents of the two facets to have
opposite signs.  Similarly, some orthorhombicity is necessary for
the critical current not to be zero on the grain boundaries of
Fig.~\protect\ref{fig2}.

One measure of the degree of orthorhombicity of the
superconductivity in YBCO, is the ratio of the penetration depths,
$\lambda_a/\lambda_b$, measured along the crystallographic $a$ and
$b$ directions, in untwinned single
crystals.\protect\cite{bas95,sun95}  This ratio is approximately
1.5 to 2.0 in high quality crystals \protect\cite{bas95} thus
indicating a considerable degree of superconducting
orthorhombicity.  The result $\lambda_a/\lambda_b\sim 1$ obtained
in an earlier measurement \protect\cite{sch90} has been attributed
\protect\cite{bas95} to the presence of disorder in the CuO chains.
Since the degree of orthorhombicity appears to depend on crystal
quality, and since the orientation $\delta$ of the grain boundary
(see Fig.~\protect\ref{fig1}) at which
the critical Josephson current vanishes is also a measure of the
orthorhombicity, the crystal quality is expected to play some role
in the detailed understanding of spontaneous flux generation in
asymmetric $45^{\circ}$ grain boundaries.

\section*{acknowledgements}

I would like to thank J.R. Kirtley for a stimulating discussion
of the results of Refs.\protect\cite{man96} and
\protect\cite{mol96} and for prepublication copies of these
references, and to acknowledge the support of the
Natural Sciences and Engineering Research Council of
Canada.

\begin{figure}
\caption{Two tetragonal $d_{x^2-y^2}$ superconductors with their $a$ and
$b$ axes in the plane of the figure but oriented at $45^{\circ}$
degrees relative to each other are shown schematically. The two
superconductor are on either side of a grain boundary.  The dashed
line $oo^{\prime}$ represents an ideal straight asymmetric
$45^{\circ}$ grain boundary, whereas the solid lines $ab$ and $bc$
represent two facets of a grain boundary which are rotated by
angles of $\delta$ and $\delta^{\prime}$ relative to the ideal
case.}
\label{fig1}
\end{figure}

\begin{figure}
\caption{(a) and (b) The two ways in which twin boundaries can
end at a grain boundary.  The directions of the crystallographic
$a$ and $b$ axes in the different regions are indicated by the
rectangles.}
\label{fig2}
\end{figure}

\begin{figure}
\caption{(a) The dynamical potential $V(\gamma )$ appropriate for the
description of a Josephson vortex.  (b) Qualitative
behavior of the gauge invariant phase
$\gamma $ as a function of the distance $y$ along the grain boundary for
the Josephson vortex.}
\label{fig3}
\end{figure}

\begin{figure}
\caption{(a) The dynamical potential $V(\gamma )$ appropriate for the
description of a vortex containing a flux $\frac{1}{2} \Phi_0$. (b)
The gauge invariant phase $\gamma $ for this vortex.  The origin $y=0$ is
the location of a twin boundary of the type shown in
Fig.~\protect\ref{fig2}(a).}
\label{fig4}
\end{figure}

\begin{figure}
\caption{The dynamical potential describing a Josephson vortex
pinned by a pair of neighboring twin boundaries. The changes in
slope of the potential at $b$ and $c$ are due to the twin
boundaries.}
\label{fig5}
\end{figure}

\begin{figure}
\caption{Dynamical potential for which twin boundaries are
encountered at points $b$ and $f$.}
\label{fig6}
\end{figure}
\end{document}